\documentclass[letterpaper,pra,showpacs,showkeys,10pt,twoside,byrevtex,nofootinbib, superscriptaddress, amsfonts,amsmath]{revtex4-1}
 \usepackage{graphicx}
  \usepackage{amssymb}
  \usepackage{color}
  \begin{document}
\title{Quantizing the Vector Potential Reveals Alternative Views of the  Magnetic Aharonov-Bohm Phase Shift}
\author{Philip Pearle}
\email{ppearle@hamilton.edu}
\affiliation{Emeritus, Department of Physics, Hamilton College, Clinton, NY  13323}
\author{Anthony Rizzi}
\email{arizzi@iapweb.org}
\affiliation{Institute for Advanced Physics, PO Box 15030, Baton Rouge, Louisiana, 70895}
\date{\today}
\begin{abstract} {We give a complete quantum analysis of the Aharonov- Bohm (A-B) magnetic phase shift involving three entities, the electron, the charges constituting the solenoid current, and the vector potential. The usual calculation supposes that the solenoid's vector potential may be well- approximated as classical. The A-B shift is then acquired by the quantized electron moving in this vector potential. Recently, Vaidman presented a semi-classical calculation, later confirmed by a fully quantum calculation of Pearle and Rizzi, where it is supposed that the electron's vector potential may be well-approximated as classical. The A-B shift is then acquired by the quantized solenoid charges moving in this vector potential. Here we present a third calculation, which supposes that the electron and solenoid currents may be well-approximated as classical sources. The A-B phase shift is then shown to be acquired by the quantized vector potential. We next show these are three equivalent alternative ways of calculating the A-B shift. We consider the exact problem where all three entities are quantized. We approximate the wave function as the product of three wave functions, a vector potential wave function, an electron wave function and a solenoid wave function. We apply the variational principle for the exact Schrodinger equation to this approximate form of solution. This leads to three Schrodinger equations, one each for vector potential, electron and solenoid, each with classical sources for the other two entities. However, each Schrodinger equation contains an additional real c-number term, the time derivative of an extra phase. We show that these extra phases are such that the net phase of the total wave function produces the A-B shift.  Since none of the three entities requires different treatment from any of the others, this leads to three alternative views of the physical cause of the A-B magnetic effect.}
 \end{abstract}
\maketitle
\section{Introduction.}
The exact problem for the magnetic Aharonov-Bohm (A-B) effect involves three quantized entities, the electron, the solenoid charges and the vector potential.  The Schr\"odinger equation has the form
\begin{equation}\label{1}
i\frac{d}{dt}|\psi,t\rangle=\bigg[\hat H_{el}+\hat H_{sol}+\hat H_{A}-\int d{\bf x}[\hat{\bf J}_{el}({\bf x})+\hat{\bf J}_{sol}({\bf x})]\cdot\hat {\bf A}({\bf x})\bigg]|\psi,t\rangle.
\end{equation}
\noindent $\hat H_{el}$ and $\hat H_{sol}$  are Hamiltonians for the electron and solenoid charges.  In addition to the kinetic energy terms,  if the particles move in tubes or start at rest and are accelerated to some 
final speed, this will include potentials was well. $\hat H_{A}$ is the Hamiltonian for the free quantum vector potential field.  
 $\hat{\bf J}_{el}({\bf x}), \hat{\bf J}_{sol}({\bf x})$ are the current operators of the electron and solenoid charges respectively, and $\hat{\bf A}({\bf x})$ is the vector potential operator. 
 
 One would expect that the phase shift calculated thereby would be 
the usual A-B expression if, for no other reason, that experiment shows this result\cite{Ton}.  However, this exact problem has not been solved.

What can be solved are three truncated problems, detailed in the following three paragraphs.  In each of these problems, in the interaction term, two of these entities are considered classical, and one is considered quantum. 

In the standard treatment, based upon the original work by Aharonov and Bohm\cite{AB, FR, AR},  the choice is made to approximate 
the solenoid current  as  classical, $\hat{\bf J}_{sol}({\bf x})\rightarrow {\bf J}_{sol}({\bf x})$, where ${\bf J}_{sol}({\bf x})$ is the expectation value of $\hat{\bf J}_{sol}({\bf x})$.   
The vector potential due to the solenoid is also approximated as classical, $\hat{\bf A}({\bf x})\rightarrow {\bf A}_{sol}({\bf x})$, with ${\bf J}_{sol}({\bf x})$ as source of 
${\bf A}_{sol}({\bf x})$.   Omitting the  completely classical interaction term $\int d{\bf x}{\bf J}_{sol}({\bf x})\cdot{\bf A}_{sol}({\bf x})$, 
the  resulting interaction term is therefore $\int d{\bf x}\hat{\bf J}_{el}({\bf x})\cdot{\bf A}_{sol}({\bf x})$. Thus, only the quantized electron undergoes an interaction. 
Then, the phase shift associated to the electron moving in this classical vector potential, the A-B phase shift, was found. 
 
One might cavil that such a vector potential can hardy be considered classical since, unlike anything classical, it is force-free yet has a physical effect. 
 But, classicizing the vector potential is accepted and its effect on the quantized electron is taken to illustrate  
one of the marvelous distinctions between classical and quantum physics.

However, recently Vaidman\cite{Vaidman} chose to approximate the electron current as classical,  $\hat{\bf J}_{el}({\bf x})\rightarrow {\bf J}_{el}({\bf x},t)$, where ${\bf J}_{el}({\bf x},t)$ is the expectation value of $\hat{\bf J}_{el}({\bf x})$. Moreover, the vector potential due to the electron is also approximated as classical, $\hat{\bf A}({\bf x})\rightarrow {\bf A}_{el}({\bf x},t)$, 
with ${\bf J}_{el}({\bf x},t)$ as source of ${\bf A}_{el}({\bf x},t)$.  Omitting the  completely classical interaction term $\int d{\bf x}{\bf J}_{el}({\bf x},t)\cdot{\bf A}_{el}({\bf x},t)$, 
the resulting interaction term is therefore $\int d{\bf x}\hat{\bf J}_{sol}({\bf x})\cdot{\bf A}_{el}({\bf x},t)$. Thus, only the quantized solenoid charges undergo an interaction. Vaidman showed  by a semi-classical calculation (verified recently by a fully quantum mechanical calculation\cite{PR})  that the solenoid charges provide a phase shift identical to the usual A-B phase shift.  

In Section II of this paper,  we  choose to approximate both  the solenoid current and the electron current as classical. The resulting interaction is therefore 
$\int d{\bf x}[{\bf J}_{el}({\bf x},t)+{\bf J}_{sol}({\bf x})]\cdot\hat {\bf A}({\bf x})$.  Thus, only the quantized vector potential undergoes an interaction. We then solve for the 
 vector potential wave function.  

In Section III, we show that the latter provides a phase shift identical to the usual A-B phase shift. 

 We propose viewing these as three alternative but equally valid (although conceptually and mathematically different) ways of obtaining the same result.  
 
 However, there is the following to be considered, which seems to raise an objection to this point of view.  
 
 There are three additional  problems one might think of, in each of which only one entity is made classical in the interactions and the  remaining two are  treated as quantum mechanical.  From the successes 
where two entities are made classical, one might expect to obtain the A-B phase shift when only one entity is made classical.  
 
 Two of these problems are not readily soluble, so one cannot ascertain the phase shift for them, one where the electron and vector potential are quantized, the other where the solenoid charges and vector potential 
 are quantized.  
 
 However, the problem  can be readily solved where the vector potential is treated classically and the evolution of the quantized  electron and solenoid charges is  governed by the 
  Schr\"odinger equation:
 \begin{equation}\label{2}
i\frac{d}{dt}|\psi,t\rangle=\bigg[\hat H_{el}+\hat H_{sol}-\int d{\bf x}[\hat{\bf J}_{el}({\bf x})\cdot {\bf A}_{sol}({\bf x})+\hat{\bf J}_{sol}({\bf x})\cdot {\bf A}_{el}({\bf x},t)]\bigg]|\psi,t\rangle
\end{equation}
\noindent (omitting the self-interacting terms $\int d{\bf x}\hat{\bf J}_{el}({\bf x})\cdot {\bf A}_{el}({\bf x}),\int d{\bf x}\hat{\bf J}_{sol}({\bf x})\cdot {\bf A}_{sol}({\bf x})$ which are the same for both traverses of the electron around the solenoid,  left and right, and therefore do not contribute to the phase shift). 
 According to this equation, since the Hamiltonian is separable, both mechanisms are operating, the electron acquires the usual A-B phase shift moving in the solenoid's vector potential, and the solenoid charges acquire the A-B phase shift moving in the electron's vector potential.
Thus, the net phase shift is twice the A-B shift.  So, we see that the prescription of just letting the vector potential be classical is incorrect, at least in this case.  

That the  Schr\"odinger equation (\ref{2}) is not the correct one to use for the problem of jointly quantized electron and solenoid (with classical vector potential) was shown in reference\cite{PR}. There, a better approximation  was found, starting with the variational principle for the Schr\"odinger equation, and the A-B phase resulted (more details below).   

Here,  in Section IV, we consider the same prescription, a better approximation to the exact problem  of jointly quantized electron and solenoid and vector potential.     
We start with the variational principle for the Schr\"odinger equation, with the exact state vector replaced by the product 
$|\Psi,t\rangle\approx |\psi_{A},t\rangle |\psi_{el}, t\rangle|\psi_{sol}, t\rangle$, 
where the operator dependence of $|\psi_{A},t\rangle$ is just the vector potential, $|\psi_{el}, t\rangle$'s  dependence is just the electron operators and 
$|\psi_{sol}, t\rangle$'s dependence is just the solenoid operators. 
The variational principle produces three Schr\"odinger equations 
for the three state vectors. Each equation describes one entity interacting with classical (expectation) values of the other two entities, as well as an additional phase term.  

There is the freedom to add  phases to two state vectors and the negative of these two phases  to the other state vector without affecting the overall phase of the wave function. 
We choose to use this to remove the extra phases from the electron and solenoid Schr\"odinger equations.  
Then, \textit{except} for the extra phase term, $|\psi_{A},t\rangle$ becomes the  solution of the Schr\"odinger equation already discussed in Sections II, III, whose phase contribution to the A-B effect is, 
for the right traverse of the electron, correctly, 1/2 the A-B phase, $\equiv\Phi_{R}(T)$.  The extra phase term is shown to provide 
$-2\Phi_{R}(T)$, so the net contribution to the phase of the vector potential Schr\"odinger equation \textit{with} the extra phase is $-\Phi_{R}(T)$.

Since both the electron and solenoid Schr\"odinger equations each produce $\Phi_{R}(T)$ the net phase of the product wave function of all three  
state vectors is, correctly, $\Phi_{R}(T)$. 

It should be remarked that reference \cite{PR} considers the exact problem of the quantized 
 electron and solenoid where the vector potential is not a quantum field, but is a function of the electron and solenoid 
 position and momentum operators. The same technique used here was used there, applying the variational principal which gives the  
 exact Schr\"odinger equation to an approximate state vector which is the direct product of electron and solenoid state vectors.  
 Again,  the two resulting Schr\"odinger equations contained extra phase terms.  The net extra phase was shown to produce the negative A-B shift.  
 The wave functions for electron and solenoid, without the extra phase, each give the A-B shift, and so the net result is that the  A-B shift is obtained. 
  
Therefore, these two papers show how a good approximative treatment of the exact Schr\"odinger equation describing the interaction between quantized entities responsible for the A-B shift, either the electron and solenoid or the vector potential and electron and solenoid, results in the A-B shift.

Section V discusses some  conclusions that may be drawn from these calculations.  

We use natural units, with $c=\hbar=1$.

\section{Quantized Vector Potential With Classical Source.}
	
In this section, the Hamiltonian to be considered is:
\begin{equation}\label{3}
\hat H=\int d{\bf x}\Big[\frac{1}{2}[\hat\pi_{i}^{2}({\bf x})+{\bf \nabla}\hat A^{i}({\bf x})\cdot{\bf \nabla}\hat A^{i}({\bf x})]-J^{i}({\bf x},t)\hat A^{i}({\bf x})\Big]
\end{equation}
\noindent  where we are using the summation convention for repeated indices ($i=1,2,3$).  $J^{i}({\bf x,t})$ is a general classical current source (to be specialized in Section III, where the result is applied to the A-B situation,  to the sum of the electron current and the solenoid current), 
and $[\hat A^{i}({\bf x}), \hat\pi_{j}({\bf x'})]=i\delta_{ij}\delta({\bf x}-{\bf x}')$. 

We note that this Hamiltonian expressed classically, with the commutation relations replaced by Poisson bracket relations, gives the correct equations of motion for the vector potential with classical currents.  Thus, we have employed the standard procedure to go from a classical problem to a quantum problem. 

However, this is not the usual method of quantizing an electrodynamic problem, and requires further explication. What is unusual is that, since the Coulomb fields of the electron and solenoid pieces are not germane to the A-B problem, we just omit them.  This gives the treatment of the components of the vector potential as essentially three scalar fields, which simplifies the analysis.

Since the vector potential operator components mutually commute at all locations, we can express the wave function  in their eigenbasis. 
 For brevity of notation, we shall designate such an eigenvector as $|{\bf A}\rangle$, which satisfies 
$\hat A^{i}({\bf x})|{\bf A}\rangle=A^{i}({\bf x})|{\bf A}\rangle$, where the eigenvalues $A^{i}({\bf x})$ are different functions of ${\bf x}$ for each different eigenvector ($-\infty<A^{i}({\bf x})<\infty$).  The functional integral shall be denoted $\int DA\equiv C\prod_{{\bf x}, i}\int_{-\infty}^{\infty} d A^{i}({\bf x})$, with $C$ chosen so that $\int DA |{\bf A}\rangle\langle{\bf A}|=1$.  
 
 Of course, this is not the usual quantum electrodynamics: there is no gauge invariance and, in Appendix B where we express the state vector in terms of photons, there are longitudinally polarized photons.

 In this Section we shall solve this problem.  In Section, III we shall apply this result to an approximate solution of the  A-B problem. 
 
 \subsection{Wave Function.}

 For the classical problem,  the Poisson bracket equation of motion for the vector potential which follows from the classical version of the Hamiltonian (\ref{3}), and its general solution are: 
 \begin{subequations}
\begin{eqnarray}\label{4a}
\frac{\partial^{2}}{\partial t^{2}}A_{cl}^{i}({\bf x},t)&=&{\bf \nabla}^{2} A_{cl}^{i}({\bf x},t)+J^{i}({\bf x,t}) \\              
A_{cl}^{i}({\bf x},t)&=&\frac{1}{(2\pi)^{3}}\int d{\bf k}e^{i{\bf k}\cdot {\bf x}}[e^{-i\omega t}c^{i}({\bf k})+e^{i\omega t}c^{*i}({-\bf k})]\nonumber\\ \label{4b}
&&\qquad\qquad+\frac{1}{(2\pi)^{3}}\int d{\bf x}'\int_{0}^{t}dt'J^{i}({\bf x}',t')\int d{\bf k}e^{i{\bf k}\cdot ({\bf x}-{\bf x}')}\frac{\sin \omega (t-t')}{\omega},
\end{eqnarray}
\noindent where $\omega\equiv |{\bf k}|$,  and  the $c^{i}({\bf k})$ are arbitrary.  
 \end{subequations}

We are going to solve the quantum problem by assuming a wave function of the form:
\begin{equation}\label{5}
\langle {\bf A}|\psi_{A}, t\rangle=Ne^{-\int d{\bf x} d{\bf x}'A^{i}({\bf x}) B({\bf x}-{\bf x}',t)A^{i}({\bf x}') +i\int d{\bf x}b^{i}({\bf x},t)A^{i}({\bf x}) +ic(t)}.
\end{equation}
We insert  (\ref{5}) into  Schr\"odinger's equation with the Hamiltonian (\ref{3}).  The algebra is relegated to Appendix A, where it is shown that the solution is:

\begin{eqnarray}\label{6}
\langle {\bf A}|\psi_{A}, t\rangle&=&Ne^{-\int d{\bf x} d{\bf x}'[A^{i}({\bf x})-A_{cl}^{i}({\bf x},t) ]\frac{1}{2(2\pi)^{3}}\int d{\bf k}\omega e^{i{\bf k}\cdot ({\bf x}-{\bf x}')}[A^{i}({\bf x}')-A_{cl}^{i}({\bf x},t)]
+i\int d{\bf x}\dot A_{cl}^{i}({\bf x},t)[A^{i}({\bf x})-A_{cl}^{i}({\bf x},t) ]} \nonumber\\
&&e^{\frac{i}{2}\int d{\bf x}\dot A_{cl}^{i}({\bf x},t)A_{cl}^{i}({\bf x},t) +\frac{i}{2}\int_{0}^{t}dt'\int d{\bf x}A_{cl}^{i}({\bf x},t')J^{i}({\bf x},t')}.
\end{eqnarray}

It is worth remarking that there is no free parameter allowing adjustment of the width of $|\langle {\bf A}|\psi_{A}, t\rangle|$, unlike, e.g., the 
case for the initial state of a particle, $\langle x|\psi\rangle \sim e^{-\frac{(x-x_{0})^{2}}{\sigma^{2}}}$.   However, an adjustable width $\sigma$ appears if, instead of a point electron current source for the vector potential, the electron charge density is ``smeared" over a distance $\sigma$. In the next section, such a smearing is shown to be necessary. 

In Appendix B, it is shown that this state vector can be expressed in terms of coherent states of photons: 
 \begin{equation}\label{7}
|\psi_{A},t\rangle =e^{\int d{\bf k}\alpha^{i}({\bf k},t)a^{i\dagger}({\bf k},t )}|0\rangle e^{-\frac{1}{2}\int d{\bf k}|\alpha^{i}({\bf k},t)|^{2}}e^{\frac{i}{2}\int_{0}^{T}dt \int d{\bf x}A^{i}_{cl}({\bf x},t)J^{i}({\bf x},t)},
\end{equation}
\noindent where
\begin{eqnarray}\label{8}
\alpha^{i}({\bf k},T)&=&\frac{1}{(2\pi)^{3/2}}\int d{\bf x}e^{-i{\bf k}\cdot{\bf x}}\Bigg[\sqrt{\frac{\omega}{2}} A_{cl}^{i}({\bf x},t)+i\frac{1}{\sqrt{2\omega}} \dot A_{cl}^{i}({\bf x},t)\Bigg]                   
\end{eqnarray}
\noindent and, as usual, $\hat A^{i}({\bf x})=\frac{1}{(2\pi)^{3/2}}\int \frac{d{\bf k}}{\sqrt{2\omega}}[e^{i{\bf k}\cdot{\bf x}} a^{i}({\bf k})+e^{-i{\bf k}\cdot{\bf x}}a^{i\dagger}({\bf k})]  $.
In what follows, we shall only utilize Eq.(\ref{6}).

\section{A-B Phase of Vector Potential Wave Function.}

We now turn to apply Eq.(\ref{6}) to the special case of the A-B situation.   

The wave function for the A-B effect is the sum of two wave functions, one for the right traverse of the electron and one for the left traverse of the electron.  

In the approximation considered here, the only interaction is that of the quantized vector potential interacting with the classical electron and solenoid currents.  
These classical currents are ${\bf J}_{sol}({\bf x})$, the constant current for the solenoid, and ${\bf J}_{el}({\bf x},t)$, the current of a classical electron orbiting the solenoid with speed $u$ 
at radius $R$ in a half-circle (counterclockwise for the right traverse, clockwise for the left traverse). Then,  the wave function for each traverse is the product of  three  independently evolving wave functions, those of the vector potential, electron and solenoid.  The solenoid wave function is the same for either traverse.  The electron wave packets produces no phase of their own that differs for the two traverses.  Thus, we are considering a wave function of the form
\[ 
|\psi,t\rangle=\frac{1}{\sqrt{2}}[|\psi_{R},t\rangle+|\psi_{L},t\rangle]=\frac{1}{\sqrt{2}}|\psi_{sol},t\rangle[|\psi_{el,R},t\rangle|\psi_{A,R},t\rangle+|\psi_{el,L},t\rangle|\psi_{A,L},t\rangle],
\]
\noindent where the Schr\"odinger equation for the right traverse satisfies this approximation to Eq.(\ref{1}):
\[ 
i\frac{d}{dt}|\psi_{R},t\rangle=\bigg[\hat H_{el}+\hat H_{sol}+\hat H_{A}-\int d{\bf x}[{\bf J}_{el,R}({\bf x})+{\bf J}_{sol}({\bf x})]\cdot\hat {\bf A}({\bf x})\bigg]|\psi_{R},t\rangle
\]
(and similarly for the left traverse). 

After completion of the traverses of the packets at time $T=\pi R/u$, the two electron packets are presumed to meet the electron equivalent of a half-silvered mirror, resulting in $\frac{1}{\sqrt{2}}$ times the sum of packets emerging from one side and  $\frac{1}{\sqrt{2}}$ times the difference emerging from the other.   Since neither the electron or the solenoid contribute a phase shift, the probabilities of detecting the electron on one side or the other is $\frac{1}{2}[1\pm e^{-a(T)}\cos\Phi(T)]$, where $e^{-a(t)}e^{i\Phi(t)}= \langle \psi_{A,L},t|\psi_{A,R},t\rangle$.

  It is shown in Appendix C that the part of the wave function in the first line of Eq.(\ref{6}) produces no phase shift. 
 Thus,  the exponent  
in the second line of (\ref{6}) is responsible for the phase (the difference of which for the two trajectories gives the phase shift).  

We shall look at that phase in a moment but, first, it is also shown in Appendix C that with the classical electron current that of a point particle, then 
 there is an ultraviolet (short distance) divergence causing $a(T)=\infty$.  However, it is also shown, with the electron's charge density ``smeared" over a length the size $\sigma$ of an electron wave packet, the integral is finite, with $ a(T)\sim \frac{e^{2}}{\hbar c}\frac{u}{c}\frac{R}{\sigma}<<1$: note the tell-tale $\sigma^{-1}$ dependence.
(We may consider  $\frac{R}{\sigma}\approx 1$, which was the case in the experiment\cite{Ton} where the packets traveling through and outside the magnetized torus were the order of the torus size.)  With $e^{-a(T)}\approx 1$, there is maximum interference.

Return now to the phase in the second line of (\ref{6}) and, to be concrete, suppose we evaluate it at time $T$ at the end of the electron's right traverse:
\begin{eqnarray}\label{9}
\Phi(T)&\equiv&\frac{1}{2}\int d{\bf x}\dot A_{cl}^{i}({\bf x},T)A_{cl}^{i}({\bf x},T)+\frac{1}{2}\int_{0}^{T}dt'\int d{\bf x}A_{cl}^{i}({\bf x},t')J^{i}({\bf x},t'),\nonumber\\
&\equiv&\Phi_{1}(T)+\Phi_{2}(T).
\end{eqnarray}
\noindent 

Consider  $\Phi_{2}(T)$ first.  We substitute $A_{cl}^{i}({\bf x},t')=A_{el}^{i}({\bf x},t')+A_{sol}^{i}({\bf x})$ 
and $J^{i}({\bf x},t')=J_{el}^{i}({\bf x},t')+J_{sol}^{i}({\bf x})$, and drop terms  $A_{el}^{i}({\bf x},t')J_{el}^{i}({\bf x},t')$, $A_{sol}^{i}({\bf x},t')J_{sol}^{i}({\bf x})$ which are the scalar product of two terms with the same subscript, 
since they are the same for both traverses of the electron\footnote {This is obvious for ${\bf A}_{sol}\cdot {\bf J}_{sol}$. For the electron, the two trajectories 
are exchanged under a rotation of $180^{\circ}$ about the $y$-axis, so  
$\int d{\bf x}A_{el}^{i}({\bf x},t')J_{el}^{i}({\bf x},t')$ is unchanged.}: 
  \begin{equation}\label{10}
\Phi_{2}(T)\equiv\Phi_{21}(T) +\Phi_{22}(T)     \equiv\frac{1}{2}\int d{\bf x}\int_{0}^{T}dt'\Big[J_{el}^{i}({\bf x},t')A_{sol}^{i}({\bf x})+A_{el}^{i}({\bf x},t')J_{sol}^{i}({\bf x})\Big].
\end{equation}
\noindent $\Phi_{2}(T)$ is thus 1/2 the sum of a
 phase due to the electron moving in the vector potential of the solenoid 
and a phase due to the solenoid moving in the vector potential of the electron.

The integral in the first term is the well-known contribution, to the usual A-B phase shift $\Phi_{AB}$, of the right traverse of the electron:
  \begin{equation}\label{11}
2\Phi_{21}(T)=\int d{\bf x}\int_{0}^{T}dt'J_{el}^{i}({\bf x},t')A_{sol}^{i}({\bf x})=\int_{0}^{T}dt'e{\bf u}_{el}(t')\cdot A_{sol}^{i}({\bf x}_{el}(t'))=
\int_{0}^{{\bf x}_{el}(T)}ed{\bf x}_{el}\cdot A_{sol}^{i}({\bf x}_{el})=\frac{1}{2}\Phi_{AB},
\end{equation}
 \noindent where we have used $J_{el}^{i}({\bf x},t')=e{\bf u}_{el}(t')\delta({\bf x}-{\bf x}_{el}(t'))$ and ${\bf u}_{el}(t')dt'=d{\bf x}_{el}$.
 
The integral in the second term,  $2\Phi_{22}(T)$, was shown in \cite{PR} to be 
the phase contributed by the solenoid charges moving under the vector potential of the right traverse of the electron. 
 In the non-relativistic approximation, where  $\nabla^{2}A_{el}^{i}({\bf x},t)\approx -J_{el}^{i}({\bf x},t)$, 
this is equal to the right traverse's contribution to the A-B phase shift:
 \begin{eqnarray}\label{12}
 2\Phi_{22}(T)&=&\int d{\bf x}\int_{0}^{T}dt'A_{el}^{i}({\bf x},t')J_{sol}^{i}({\bf x})=\int d{\bf x}\int_{0}^{T}dt'A_{el}^{i}({\bf x},t')(-)\nabla^{2}A_{sol}^{i}({\bf x})\nonumber\\
 &=&
 \int d{\bf x}\int_{0}^{T}dt'(-)\nabla^{2}A_{el}^{i}({\bf x},t')A_{sol}^{i}({\bf x})=\int d{\bf x}\int_{0}^{T}dt'J_{el}^{i}({\bf x},t')A_{sol}^{i}({\bf x})= 2\Phi_{21}(T)=\frac{1}{2}\Phi_{AB}.
 \end{eqnarray}
\noindent 
However, here, a relativistic calculation is being done.  Therefore, we must use $\nabla^{2}A_{el}^{i}({\bf x},t)-\ddot A_{el}^{i}({\bf x},t)=-J_{el}^{i}({\bf x},t)$.  
Therefore,  to get the same result as (\ref{12})  requires the inclusion of $\Phi_{1}(T)$.
We now show that  $2\Phi_{22}(T)$ and $2\Phi_{1}(T)$, combine to give $\frac{1}{2}\Phi_{AB}$.  

The first three steps of (\ref{12}) are valid: 
\begin{eqnarray}\label{13}
2\Phi_{22}(T)&\equiv &\int d{\bf x}\int_{0}^{T}dt'A_{el}^{i}({\bf x},t')J_{sol}^{i}({\bf x})=-\int d{\bf x}\int_{0}^{T}dt'A_{el}^{i}({\bf x},t')\nabla^{2}A_{sol}^{i}({\bf x})\nonumber\\
&=&-\int d{\bf x}\int_{0}^{T}dt'\nabla^{2}A_{el}^{i}({\bf x},t')A_{sol}^{i}({\bf x})
\end{eqnarray}

Now, $\Phi_{1}(T)$, although evaluated at time $T$, actually depends upon the time interval $(0,T)$.  We note that 
$\dot A_{cl}^{i}({\bf x},t)=d[ A_{sol}^{i}({\bf x})+A_{el}^{i}({\bf x},t)] =\dot A_{el}^{i}({\bf x},t)$.  When the electron accelerates, it begins to radiate, spheres of vector potential 
move out with the speed of light from each point on its trajectory,  so $\dot A_{el}^{i}({\bf x},t)$ is non-zero throughout the radiated region. 

Disregarding $\int d{\bf x}\dot A_{el}^{i}({\bf x},T)A_{el}^{i}({\bf x})$ as it is the same for both traverses of the electron, we have
\begin{eqnarray}\label{14}
2\Phi_{1}(T)&=&\int d{\bf x}\dot A_{el}^{i}({\bf x},T)A_{sol}^{i}({\bf x})=\int d{\bf x}\int_{0}^{T}dt'\frac{\partial}{\partial t'}[\dot A_{el}^{i}({\bf x},t')A_{sol}^{i}({\bf x})]\nonumber\\
&=&
\int d{\bf x}\int_{0}^{T}dt'\Bigg[\frac{\partial^{2}}{\partial t'^{2}}A_{el}^{i}({\bf x},t')\Bigg]A_{sol}^{i}({\bf x})
\end{eqnarray}
\noindent (we have assumed that the electron is at rest at time $0+$, and accelerates immediately thereafter, so the boundary value  $\dot A_{el}^{i}({\bf x},0)=0$).
Therefore, 
\begin{eqnarray}\label{15}
2\Phi_{1}(T)+2\Phi_{22}(T)&=&\int d{\bf x}\int_{0}^{T}dt'\Bigg[\frac{\partial^{2}}{\partial t'^{2}}A_{el}^{i}({\bf x},t')-\nabla^{2}A_{el}^{i}({\bf x},t')\Bigg]A_{sol}^{i}({\bf x})\nonumber\\
&=&\int d{\bf x}\int_{0}^{T}dt'J_{el}^{i}({\bf x},t')A_{sol}^{i}({\bf x})=\frac{1}{2}\Phi_{AB}.
\end{eqnarray}
\noindent Thus, the electron's right traverse produces the phase $\Phi(T)=\Phi_{21}(T)+(\Phi_{1}(T)+\Phi_{22}(T))=\frac{1}{4}\Phi_{AB}+\frac{1}{4}\Phi_{AB}=\frac{1}{2}\Phi_{AB}$,
the left traverse produces the phase $-\frac{1}{2}\Phi_{AB}$, which is to be subtracted, giving the total  $\Phi_{AB}$.

This concludes our demonstration that 
 the A-B phase shift is obtained from the phase of the wave function describing  the quantized vector potential 
with classical electron and solenoid current sources. 

\section{Approximate Solution For Exact A-B Problem.}

For our next enterprise, we turn to approximating the solution of the exact problem, Schr\"odinger's equation (\ref{1}).  

Consider the variational problem 
$\delta I=0$, where
\begin{equation}\label{16}
I\equiv\int_{0}^{T} dt\Big[\langle \psi,t|i\frac{d}{dt}|\psi, t\rangle-\langle \psi,t|\hat H|\psi, t\rangle\Big]. 
\end{equation}
\noindent The variation with respect to $\langle \psi,t|$ gives the Schr\"odinger equation, and the independent variation of $|\psi, t\rangle$ gives its Hermitian conjugate. 

We consider a situation where  the electron and solenoid particles 
are well-localized.  In that circumstance, for either the left or right traverse of the electron around the solenoid, we approximate the actual state vector by a direct product of all three involved entities: 
$|\Psi,t\rangle\approx |\psi_{A},t\rangle|\psi_{el}, t\rangle|\psi_{sol}, t\rangle$.  Also, we consider that well-localized means the expectation value of the 
current operators are well-approximated by the corresponding classical currents: 
$\langle \psi_{el},t|\hat{\bf J}_{el}({\bf x})|\psi_{el}, t\rangle\approx {\bf J}_{el}({\bf x},t)$,  $\langle \psi_{sol},t|\hat{\bf J}_{sol}({\bf x})|\psi_{el}, t\rangle\approx {\bf J}_{sol}({\bf x},t)$, 
${\bf J}({\bf x},t)\equiv {\bf J}_{el}({\bf x},t)+{\bf J}_{sol}({\bf x},t)$.

\subsection{Schr\"odinger Equations.}

Upon substituting $|\Psi,t\rangle\approx |\psi_{A},t\rangle|\psi_{el}, t\rangle|\psi_{sol}, t\rangle$  into (\ref{16}) with $\hat H$ given by (\ref{1}) and varying 
$\langle \psi_{A},t|$, $\langle \psi_{el}, t|$ and $\langle \psi_{sol}, t|$ independently, we obtain three coupled equations:
\begin{subequations}
\begin{eqnarray}\
&&i\frac{d}{dt}|\psi_{A},t\rangle+\Bigg\{\langle \psi_{el}, t|\Big[i\frac{d}{dt}-\hat H_{el}\Big]|\psi_{el}, t\rangle
+\langle \psi_{sol}, t|\Big[i\frac{d}{dt}-\hat H_{sol}\Big]|\psi_{sol}, t\rangle\Bigg\}|\psi_{A},t\rangle\nonumber\\
&&\qquad\qquad=\hat H_{A}|\psi_{A},t\rangle
-\int d{\bf x}[\langle \psi_{el}, t|\hat {\bf J}_{el}({\bf x})|\psi_{el}, t\rangle+ \langle \psi_{sol}, t|\hat {\bf J}_{sol}({\bf x})|\psi_{sol}, t\rangle]  \cdot \hat{\bf A}({\bf x},t)|\psi_{A},t\rangle
 \approx [\hat H_{A}- {\bf J}({\bf x},t) \cdot \hat{\bf A}({\bf x},t)]|\psi_{A},t\rangle,\nonumber\\
  \label{17a}\\      
&&i\frac{d}{dt}|\psi_{el},t\rangle+\Bigg\{\langle \psi_{A}, t|\Big[i\frac{d}{dt}-\hat H_{A}\Big]|\psi_{A}, t\rangle 
+\langle \psi_{sol}, t|\Big[i\frac{d}{dt}-\hat H_{sol}\Big]|\psi_{sol}, t\rangle\nonumber\\
&&\qquad\qquad\qquad\qquad\qquad\qquad\qquad\qquad\qquad\qquad+\int d{\bf x}\langle \psi_{A}, t|\hat {\bf A}({\bf x},t)|\psi_{A}, t\rangle \cdot \langle\psi_{sol},t|\hat{\bf J}_{sol}({\bf x})|\psi_{sol},t\rangle\Bigg\}|\psi_{el},t\rangle
\nonumber\\
&&\qquad\qquad=\hat H_{el}|\psi_{el},t\rangle
-\int d{\bf x}\langle \psi_{A}, t|\hat {\bf A}({\bf x},t)|\psi_{A}, t\rangle \cdot \hat{\bf J}_{el}({\bf x})|\psi_{el},t\rangle=
\hat H_{el}|\psi_{el},t\rangle
-\int d{\bf x}{\bf A}_{cl}({\bf x},t) \cdot \hat{\bf J}_{el}({\bf x})|\psi_{el},t\rangle,\label{17b}\\
&&i\frac{d}{dt}|\psi_{sol},t\rangle+\Bigg\{\langle \psi_{A}, t|\Big[i\frac{d}{dt}-\hat H_{A}\Big]|\psi_{A}, t\rangle 
+\langle \psi_{el}, t|\Big[i\frac{d}{dt}-\hat H_{el}\Big]|\psi_{el}, t\rangle \nonumber\\
&&\qquad\qquad\qquad\qquad\qquad\qquad\qquad\qquad\qquad\qquad+
\int d{\bf x}\langle \psi_{A}, t|\hat {\bf A}({\bf x},t)|\psi_{A}, t\rangle \cdot \langle\psi_{el},t|\hat{\bf J}_{el}({\bf x})|\psi_{el},t\rangle\Bigg\}|\psi_{sol},t\rangle
\nonumber\\
&&\qquad\qquad=\hat H_{sol}|\psi_{sol},t\rangle
-\int d{\bf x}\langle \psi_{A}, t|\hat {\bf A}({\bf x},t)|\psi_{A}, t\rangle \cdot \hat{\bf J}_{sol}({\bf x})|\psi_{sol},t\rangle=
\hat H_{sol}|\psi_{sol},t\rangle
-\int d{\bf x}{\bf A}_{cl}({\bf x},t) \cdot \hat{\bf J}_{sol}({\bf x})|\psi_{sol},t\rangle.
\label{17c}
\end{eqnarray}
\end{subequations}
\noindent We have anticipated that the equations are of Hamiltonian form so the  time evolution is unitary, and we may take $\langle \psi_{A},t|\psi_{A},t\rangle=\langle \psi_{el, sol}, t|\psi_{el, sol}, t\rangle=1$. We have used the fact that  $\langle \psi_{A}, t|\hat {\bf A}({\bf x},t)|\psi_{A}, t\rangle={\bf A}_{cl}({\bf x},t)$  for this state vector,
in the last line of Eqs.(\ref{17b},\ref{17c}).  

Eq.(\ref{17a}) is the equation we have considered, in Sections II, III, for the quantized vector potential with classical electron and solenoid sources \textit{except} 
for the additional term in the curly brackets.  That term is real (take the complex conjugate and use  $d\langle\psi_{el},t| \psi_{el},t\rangle/dt=
d\langle\psi_{sol},t| \psi_{sol},t\rangle/dt           =0$), 
so it is the time derivative of a phase.  Without that term, we have found the phase of the vector potential wave function, We have seen that, in the magnetic A-B situation, for e.g., the right traverse of the electron, it gives half the A-B phase shift.

Eq.(\ref{17b}) is the equation for the quantized electron with a classical vector potential due to the solenoid (the classical vector 
potential part due to the electron itself may be ignored, as this self-interaction phase contribution  is the same for left and right traverses of the electron) \textit{except} 
for the additional phase term in the curly brackets.  Without that term, the phase of the electron wave function, for the right traverse of the electron, 
gives of course half the A-B phase shift. 

Eq.(\ref{17c}) is the equation for the quantized solenoid with a classical vector potential due to the electron (the classical vector 
potential part due to the solenoid itself may be ignored, as this self-interaction phase contribution is the same for left and right traverses of the electron) \textit{except} 
for the additional phase term in the curly brackets.  Without that term, the phase of the solenoid wave function, for the right traverse of the electron, 
was shown in \cite {PR}  and in the previous section, that it gives half the A-B phase shift.

\subsection{The Extra Phase.}

And now, we note  that adding a phase factor to each state vector does not change the phase of the overall 
state vector provided the phases sum to zero. We shall use this to eliminate the extra phase term from the electron and solenoid Schr\"odinger equations (although  
we could have made a different choice). 

Given any solution for the three state vectors, write 
$|\psi_{el},t\rangle=|\psi'_{el},t\rangle e^{i\phi_{el}(t)}$, 
$|\psi_{sol},t\rangle=|\psi'_{sol},t\rangle e^{i\phi_{sol}(t)}$, $|\psi_{A},t\rangle=|\psi'_{A},t\rangle e^{-i[\phi_{el}(t)+\phi_{sol}(t)]}$, where 
$\dot\phi_{el}(t)$ is the curly bracket term in the electron's Schr\"odinger equation and $\dot\phi_{sol}(t)$ is the curly bracket term in the solenoid's Schr\"odinger equation.
Then, $|\Psi,t\rangle= |\psi_{A},t\rangle|\psi_{el}, t\rangle|\psi_{sol}, t\rangle=|\psi'_{A},t\rangle|\psi'_{el}, t\rangle|\psi'_{sol}, t\rangle$ and  we get (note that in (\ref{17a}), 
when the substitution of unprimed for primed state vectors is made and the time derivatives of the phase factors are taken, $\dot\phi_{el}(t),\dot\phi_{sol}(t)$  produce no net contribution):
\begin{subequations}
\begin{eqnarray}\
&&i\frac{d}{dt}|\psi'_{A},t\rangle+\Bigg\{\langle \psi'_{el}, t|\Big[i\frac{d}{dt}-\hat H_{el}\Big]|\psi'_{el}, t\rangle
+\langle \psi'_{sol}, t|\Big[i\frac{d}{dt}-\hat H_{sol}\Big]|\psi'_{sol}, t\rangle\Bigg\}\psi'_{A},t\rangle
= [\hat H_{A}- {\bf J}({\bf x},t) \cdot \hat{\bf A}({\bf x},t)]|\psi'_{A},t\rangle,\nonumber\\
  \label{18a}\\      
&&i\frac{d}{dt}|\psi'_{el},t\rangle=
\hat H_{el}|\psi'_{el},t\rangle
-\int d{\bf x}{\bf A}_{cl}({\bf x},t) \cdot \hat{\bf J}_{el}({\bf x})|\psi'_{el},t\rangle.
\label{18b}\\
&&i\frac{d}{dt}|\psi'_{sol},t\rangle=
\hat H_{sol}|\psi'_{sol},t\rangle
-\int d{\bf x}{\bf A}_{cl}({\bf x},t) \cdot \hat{\bf J}_{sol}({\bf x})|\psi'_{sol},t\rangle.
\label{18c}
\end{eqnarray}
\end{subequations}

The extra phase terms in (\ref{18a}) are obtained from (\ref{18b}), (\ref{18c}) by taking the expectation values:
\begin{subequations}
\begin{eqnarray}\  
\langle\psi'_{el},t|\Big[ i\frac{d}{dt}-\hat H_{el}\Big]|\psi'_{el},t\rangle&=&
-\langle\psi'_{el},t|\int d{\bf x}{\bf A}_{cl}({\bf x},t) \cdot \hat{\bf J}_{el}({\bf x})|\psi'_{el},t\rangle=-\int d{\bf x}{\bf A}_{cl}({\bf x},t)\cdot {\bf J}_{el}({\bf x},t),
\label{19a}\\
\langle\psi'_{sol},t|\Big[ i\frac{d}{dt}-\hat H_{sol}\Big]|\psi'_{sol},t\rangle&=&
-\langle\psi'_{sol},t|\int d{\bf x}{\bf A}_{cl}({\bf x},t) \cdot \hat{\bf J}_{sol}({\bf x})|\psi'_{sol},t\rangle=-\int d{\bf x}{\bf A}_{cl}({\bf x},t)\cdot {\bf J}_{sol}({\bf x},t).
\label{19b}
\end{eqnarray}
\end{subequations}
\noindent The right hand sides of (\ref{19a}), (\ref{19b}) are the same, e.g., for the electron's right traverse, both equal to -$\dot\Phi(t)$, where $\Phi(T)=\frac{1}{2}\Phi_{AB}$.   
  Thus, (\ref{18a}) becomes
\begin{equation}  \label{20}
i\frac{d}{dt}|\psi'_{A},t\rangle-2\dot\Phi(t)
= [\hat H_{A}- {\bf J}({\bf x},t) \cdot \hat{\bf A}({\bf x},t)]|\psi'_{A},t\rangle.
\end{equation}

For its right traverse (a similar result holds for the left traverse), the phase contribution of the electron wave function satisfying (\ref{18b}) is $e^{i\Phi(t)}$, as is the phase contribution of the solenoid 
 wave function satisfying (\ref{18c}). The phase contribution of the vector potential wave function satisfying (\ref{18a}) \textit{without} the extra phase term is also 
 $e^{i\Phi(t)}$, but with the extra phase terms it is $e^{-i\Phi(t)}$,  Therefore, the combined phase is $e^{i\Phi(t)}$, and at time $T$ this is the correct phase contribution 
 to the A-B phase shift,
  $e^{i\frac{1}{2}\Phi_{AB}}$.
 
\section{Concluding Remarks.}

A well-trained physicist is supposed to know intuitively how to split the world into classical and quantum, 
in order to to do theoretical analysis, of the truly quantum acting in a classical background.  The magnetic A-B effect 
appears to provide an example where one does not need to be well-trained.  For the interaction, the split of electron, solenoid and vector potential contributions 
into two classical and one quantum may be made any way, and one gets the right answer.  

A contribution of this paper has been to 
show that the quantized vector potential is an equal partner in this troika.  

Why this works isn't intuitively clear.  Reading from left to right, if asked to put in order what is most quantum (least classical) to most classical (least quantum), 
most people would make the list: electron, vector potential, solenoid. Then, why should it be that the current of the most quantum thing, the electron, can blithely be made classical in  the interaction in two of the three calculations?    

This paper provides a cautionary tale. Since the whole world  is quantum, the only sure thing is to make all three entities quantum. Then, 
one can try to proceed from there to make justifiable approximations. 

Surely, one would think, if it apparently is a good approximation (since one gets the right answer) to 
make any two things  in the interaction classical, it surely should work just fine, even be a better approximation, to make just one of the two things classical. 
But, we have seen that is not the case.   
Just replacing the vector potential operator by its classical counterpart in the interaction gives the wrong answer. 

What goes wrong with this naive approach is that it amounts to simply adding the phase shifts. 
This arises because the Hamiltonian, as shown in (\ref{2}), is then separable, and thus the wave function is simply the product of the two wave functions for the 
separate parts of the compound system.

 In our companion paper\cite{PR}, we show that a more careful approximation, while not employing the quantized vector potential field, 
is to characterize the interaction using the vector potential expressed as a function of the position and momentum operators associated with the electron and solenoid particles.  
Then, while the wave function is again approximated as the product of the electron and solenoid wave functions, a variational approximation 
just like the one used in this paper, shows that an additional phase shift must be included for the system as a whole, and 
this is essential to reconciling the results.

A second contribution of this paper has been to show, similarly, how  the magnetic A-B phase shift (highly similar considerations occur for the electric A-B effect\cite{PR}) 
arises when all three quantizeable entities are treated on an equal basis.  It is clear, within the framework of our discussion,  that 
there is no way to prefer one split of the world over another, to prefer the notion that the phase shift is due to the electron's motion in the solenoid's vector potential, 
over that it is due to  the solenoid particles' motion in the electron's vector potential, over that it is due to the vector potential evolving governed by the sources of electron and solenoid currents.  

Presumably the exact solution where all three entities are quantized would give the A-B shift.    
 One surmises it would not be possible to attribute it to any one entity, just as is the case here, with 
the approximation to the exact solution. 
 
This example may be considered to illustrate the holism of quantum theory's description of nature. 

\appendix
\section{Calculation of Wave Function for Vector Potential  With a Classical Current Source.}

We start with the wave function (\ref{5}),

\begin{equation}\label{A1}
\langle {\bf A}|\psi_{A}, t\rangle=Ne^{-\int d{\bf x} d{\bf x}'A^{i}({\bf x}) B({\bf x}-{\bf x}',t)A^{i}({\bf x}') +i\int d{\bf x}b^{i}({\bf x},t)A^{i}({\bf x}) +ic(t)}.
\end{equation}
\noindent where $B$ is symmetrical in its argument, $B({\bf x}-{\bf x}',t)=B({\bf x}'-{\bf x},t)$. 

We insert it into the Schr\"odinger equation with the Hamiltonian (\ref{3}): 
\begin{equation}\label{A2}
i\frac{d}{dt}\langle {\bf A}|\psi_{A}, t\rangle=\int d{\bf x}\Bigg[\frac{1}{2}\Big[-\frac{\delta^{2}}{\delta A^{i2}({\bf x})}+
{\bf \nabla}A^{i}({\bf x})\cdot{\bf \nabla}A^{i}({\bf x})\Big]-J^{i}({\bf x},t)A^{i}({\bf x})\Bigg]\langle {\bf A}|\psi_{A}, t\rangle.
\end{equation}
\noindent Evaluating both sides of (\ref{A2}), and dividing by $\langle {\bf A}|\psi, t\rangle$, we have:
\begin{eqnarray}\label{A3}
&&-\int d{\bf x} d{\bf x}'A^{i}({\bf x})i\frac{\partial}{\partial t} B({\bf x}-{\bf x}',t)A^{i}({\bf x}') -\int d{\bf x}\frac{\partial}{\partial t} b^{i}({\bf x},t)A^{i}({\bf x}) -\frac{\partial}{\partial t} c(t)\nonumber\\
&&=\int d{\bf x} \Bigg[\frac{1}{2}\Big[-(-2\int d{\bf x}'B({\bf x}-{\bf x}',t)A^{i}({\bf x}')+ib^{i}({\bf x},t))^{2}+2B(0,t)\Big]+\frac{1}{2}{\bf \nabla}A^{i}({\bf x})\cdot{\bf \nabla}A^{i}({\bf x})\Big]-J^{i}({\bf x},t)A^{i}({\bf x})\Bigg]\nonumber\\
&&=-2\int d{\bf x}_{1}\int d{\bf x}\int d{\bf x}'B({\bf x}_{1}-{\bf x},t)B({\bf x}_{1}-{\bf x}',t)A^{i}({\bf x})A^{i}({\bf x}')+\frac{1}{2}\int d{\bf x}\int d{\bf x}'\delta({\bf x}-{\bf x}'){\bf \nabla}A^{i}({\bf x})\cdot{\bf \nabla}A^{i}({\bf x}')\nonumber\\
&&\qquad\qquad \qquad  +2i\int d{\bf x}\int d{\bf x}'B({\bf x}-{\bf x}',t)A^{i}({\bf x}') b^{i}({\bf x},t)-\int d{\bf x}J^{i}({\bf x},t)A^{i}({\bf x})+\frac{1}{2}\int d{\bf x}b^{i2}({\bf x},t).
\end{eqnarray}
\noindent In the last step, we have dropped the phase $\int d{\bf x}B(0,t)$.

 Vanishing of the coefficients of $A^{i}({\bf x})A^{i}({\bf x}'), A^{i}({\bf x})$ and $1$ implies the three conditions
 \begin{subequations}
\begin{eqnarray}\label{A4a}
-i\dot B({\bf x}-{\bf x}',t)&=&-2\int d{\bf x}_{1} B({\bf x}-{\bf x}_{1},t)B({\bf x}'-{\bf x}_{1},t)+\frac{1}{2}{\bf \nabla}_{\bf x}\cdot {\bf \nabla}_{{\bf x}'}\delta({\bf x}-{\bf x}')\\\label{A4b}
-\dot b^{i}({\bf x},t)&=& 2i\int d{\bf x}' B({\bf x}-{\bf x}',t)b^{i}({\bf x}',t)-J^{i}({\bf x},t)\\\label{A4c}
-\dot c&=&\frac{1}{2}\int d{\bf x}' b^{i2}({\bf x}',t).
\end{eqnarray}
\end{subequations}

Eq.(\ref{A4a}) is a generalized form of the Riccati equation, and the following general solution may readily  be verified:
\begin{eqnarray}\label{A5}
B({\bf x}-{\bf x}',t)&=&\frac{1}{2(2\pi)^{3}}\int d{\bf k}\omega e^{i{\bf k}\cdot ({\bf x}-{\bf x}')}\frac{-f({\bf k})e^{-i\omega t}+e^{i\omega t}}{f({\bf k})e^{-i\omega t}+e^{i\omega t}}\nonumber\\
&=&\frac{1}{2(2\pi)^{3}}\int d{\bf k}\omega e^{i{\bf k}\cdot ({\bf x}-{\bf x}')}\frac{1-|f({\bf k})|^{2}-f({\bf k})e^{-i2\omega t}+f^{*}({\bf k})e^{2i\omega t}}{|f({\bf k})e^{-i\omega t}+e^{i\omega t}|^{2}}.
\end{eqnarray}
\noindent $f({\bf k})$ is a symmetric ($f({\bf k})=f(-{\bf k})$) function (since $B({\bf x}-{\bf x}',t)$ is symmetric), but otherwise arbitrary. 
In order that $|\langle {\bf A}|\psi, t\rangle|^{2}$ be integrable, it is necessary that the quadratic form in its exponent (found from (\ref{A1}) and (\ref{A5}), and noting that 
$-f({\bf k})e^{-i2\omega t}+f^{*}({\bf k})e^{2i\omega t}$ is imaginary),
\[
\frac{1}{(2\pi)^{3}}\int d{\bf k}\omega\Bigg| \int d{\bf x}A^{i}({\bf x}) e^{i{\bf k}\cdot {\bf x}}\Bigg|^{2}\frac{1-|f({\bf k})|^{2}}{ |f({\bf k})e^{-i\omega t}+e^{i\omega t}|^{2}},
\]
be positive definite, so we must have $|f({\bf k})|<1$.  
 In order to have a 
time-translationally invariant solution, we must make the coherent state choice $f({\bf k})=0$, so we arrive at: 
\begin{equation}\label{A6}
B({\bf x}-{\bf x}')=\frac{1}{2(2\pi)^{3}}\int d{\bf k}\omega e^{i{\bf k}\cdot ({\bf x}-{\bf x}')}.
\end{equation}

 Eq.(\ref{A4b}) may now be solved, with the result:
\begin{eqnarray}\label{A7}
b^{i}({\bf x},t)&=& \frac{1}{(2\pi)^{3}}\int d{\bf k}e^{i{\bf k}\cdot {\bf x}}  e^{-i\omega t}g^{i}({\bf k})+\frac{1}{(2\pi)^{3}}\int d{\bf x}'\int_{0}^{t}dt'J^{i}({\bf x}',t')\int d{\bf k}e^{i{\bf k}\cdot ({\bf x}-{\bf x}')}e^{-i\omega(t-t')}, 
\end{eqnarray}
\noindent where  $g^{i}({\bf k})$ is an arbitrary function.  $ib^{i}({\bf x},t)$, separated into real and imaginary parts becomes:
\begin{eqnarray}\label{A8}
ib^{i}({\bf x},t)&=&\frac{i}{2(2\pi)^{3}}\int d{\bf k}e^{i{\bf k}\cdot {\bf x}}  [e^{-i\omega t}g^{i}({\bf k})-e^{i\omega t}g^{i*}(-{\bf k})]\nonumber\\
&&\qquad\qquad+\frac{1}{(2\pi)^{3}}\int d{\bf x}'\int_{0}^{t}dt'J^{i}({\bf x}',t')\int d{\bf k}e^{i{\bf k}\cdot ({\bf x}-{\bf x}')}\sin\omega(t-t')\nonumber\\
&+&\frac{i}{2(2\pi)^{3}}\int d{\bf k}e^{i{\bf k}\cdot {\bf x}}  [e^{-i\omega t}g^{i}({\bf k})+e^{i\omega t}g^{i*}(-{\bf k})]\nonumber\\
&&\qquad\qquad+\frac{i}{(2\pi)^{3}}\int d{\bf x}'\int_{0}^{t}dt'J^{i}({\bf x}',t')\int d{\bf k}e^{i{\bf k}\cdot ({\bf x}-{\bf x}')}\cos\omega(t-t')
\end{eqnarray}
This can be expressed in terms of the notation in the classical solution (\ref{4b}) by writing $c^{i}({\bf k})=\frac{i}{2\omega}g^{i}({\bf k})$, obtaining 
\begin{equation}\label{A9}
ib^{i}({\bf x},t)=2\int d{\bf x}'B({\bf x}-{\bf x}')A_{cl}^{i}({\bf x}',t)+i\dot A_{cl}^{i}({\bf x},t).
\end{equation}
\noindent  We shall shortly see the exponent in the solution is then a quadratic form in $A^{i}({\bf x})-A_{cl}^{i}({\bf x},t)$, i.e., 
the mean value of the vector potential is the classical value. 

 Eq.(\ref{A4c}) may now be written as
\begin{eqnarray}\label{A10}
ic(t)&=&\frac{-i}{2}\int_{0}^{t}dt'\int d{\bf x}b^{i2}({\bf x},t')\nonumber\\
&=&\frac{-i}{2}\int_{0}^{t}dt'\int d{\bf x}\Bigg[\dot A_{cl}^{i2}({\bf x},t')-4\Big[\int d{\bf x}'B({\bf x}-{\bf x}')A_{cl}^{i}({\bf x}',t')\Big]^{2}\Bigg]\nonumber\\
&&\qquad\qquad-2\int_{0}^{t}dt'\int d{\bf x}d{\bf x}'\dot A_{cl}^{i}({\bf x},t')B({\bf x}-{\bf x}')A_{cl}^{i}({\bf x}',t')
\end{eqnarray}

The last (real) term of (\ref{A10}) may be written as
\begin{equation}\label{A11}
-\int_{0}^{t}dt'\frac{d}{dt'}\int d{\bf x}d{\bf x}'\ A_{cl}^{i}({\bf x},t')B({\bf x}-{\bf x}')A_{cl}^{i}({\bf x}',t')=-\int d{\bf x}d{\bf x}' A_{cl}^{i}({\bf x},t)B({\bf x}-{\bf x}')A_{cl}^{i}({\bf x}',t)+C. 
\end{equation}
\noindent  The constant term $e^{C}$ may be absorbed into the normalization constant $N$, and so removed from consideration.  

In the second (imaginary) term of Eq.(\ref{A10}), we note, using  (\ref{A6}) (or (\ref{A4a}) since $\dot B=0$), that
\begin{equation}\label{A12}
\int d{\bf x}_{1}B({\bf x}_{1}-{\bf x})B({\bf x}_{1}-{\bf x}')=\frac{1}{4(2\pi)^{3}}\int d{\bf k}\omega^{2} e^{i{\bf k}\cdot ({\bf x}-{\bf x}')}=\frac{1}{4}\nabla_{x}\cdot\nabla_{x'}\delta({\bf x}-{\bf x}').
\end{equation}

Using (\ref{A6}), (\ref{A9}), (\ref{A10}), (\ref{A11}), (\ref{A12}), and adding and subtracting $i\int d{\bf x}\dot A_{cl}^{i}({\bf x},t) A_{cl}^{i}({\bf x},t) $, we find that  the wave function (\ref{A1}) satisfying  Schr\"odinger's equation  is:
\begin{eqnarray}\label{A13}
\langle {\bf A}|\psi_{A}, t\rangle&=&Ne^{-\int d{\bf x} d{\bf x}'[A^{i}({\bf x})-A_{cl}^{i}({\bf x},t) ]\frac{1}{2(2\pi)^{3}}\int d{\bf k}\omega e^{i{\bf k}\cdot ({\bf x}-{\bf x}')}[A^{i}({\bf x}')-A_{cl}^{i}({\bf x},t)]
+i\int d{\bf x}\dot A_{cl}^{i}({\bf x},t)[A^{i}({\bf x})-A_{cl}^{i}({\bf x},t) ]} \nonumber\\
&&e^{i\int d{\bf x}\dot A_{cl}^{i}({\bf x},t)A_{cl}^{i}({\bf x},t) -\frac{i}{2}\int_{0}^{t}dt'\int d{\bf x}\big[\dot A_{cl}^{i2}({\bf x},t')-{\bf \nabla}A_{cl}^{i}({\bf x},t')\cdot{\bf \nabla}A_{cl}^{i}({\bf x},t')\big]}.
\end{eqnarray}

Write the two terms in the phase  in (\ref{A13})  as $\Phi(t)=\Phi_{1}(t)+\Phi_{2}(t)$,  where $\Phi_{2}(t)$ may be further simplified. Integrate both terms in the square bracket by parts, the first with respect to time, the second with respect to space:
\begin{eqnarray}\label{A14}
\Phi_{2}(t)&\equiv&-\frac{1}{2}\int_{0}^{t}dt'\int d{\bf x}\big[\dot A_{cl}^{i2}({\bf x},t')-{\bf \nabla}A_{cl}^{i}({\bf x},t')\cdot{\bf \nabla}A_{cl}^{i}({\bf x},t')\big]\nonumber\\
 &=&-\frac{1}{2}\int d{\bf x}\big[\dot A_{cl}^{i}({\bf x},t)A_{cl}^{i}({\bf x},t)-\dot A_{cl}^{i}({\bf x},0)A_{cl}^{i}({\bf x},0)\big]
+\frac{1}{2}\int_{0}^{t}dt'\int d{\bf x}A_{cl}^{i}\Bigg[ \frac{\partial^{2}}{\partial t'^{2}}A_{cl}^{i}({\bf x},t')- \nabla^{2}A_{cl}^{i}({\bf x},t')\Bigg]\nonumber\\
&=&-\frac{1}{2}\int d{\bf x}\dot A_{cl}^{i}({\bf x},t)A_{cl}^{i}({\bf x},t)+\frac{1}{2}\int_{0}^{t}dt'\int d{\bf x}A_{cl}^{i}({\bf x},t')J^{i}({\bf x},t').
\end{eqnarray}
\noindent In the second step we have used the dynamical equation in (\ref{4a}). We are also assuming  that the electron is initially at rest and starts moving after 
time 0, so we have the initial condition  $\dot A_{cl}^{i}({\bf x},0)=\dot A_{el}^{i}({\bf x},0)=0$.
Therefore, the solution (\ref{A13}) becomes:
\begin{eqnarray}\label{A15}
\langle {\bf A}|\psi_{A}, t\rangle&=&Ne^{-\int d{\bf x} d{\bf x}'[A^{i}({\bf x})-A_{cl}^{i}({\bf x},t) ]\frac{1}{2(2\pi)^{3}}\int d{\bf k}\omega e^{i{\bf k}\cdot ({\bf x}-{\bf x}')}[A^{i}({\bf x}')-A_{cl}^{i}({\bf x},t)]
+i\int d{\bf x}\dot A_{cl}^{i}({\bf x},t)[A^{i}({\bf x})-A_{cl}^{i}({\bf x},t) ]} \nonumber\\
&&e^{\frac{i}{2}\int d{\bf x}\dot A_{cl}^{i}({\bf x},t)A_{cl}^{i}({\bf x},t) +\frac{i}{2}\int_{0}^{t}dt'\int d{\bf x}A_{cl}^{i}({\bf x},t')J^{i}({\bf x},t')}.
\end{eqnarray}

\section{ Vector Potential Wave Function Describes a Coherent State.} \label{B}

We consider  the Hamiltonian (\ref{3}), written as :
\begin{eqnarray}\label{B1}
H&=&\int d{\bf k}\omega a^{i\dagger}({\bf k})a^{i}({\bf k})-\int d{\bf x}J^{i}({\bf x},t)\hat A^{i}({\bf x})\nonumber\\
&=&\int d{\bf k}\omega a^{i\dagger}({\bf k})a^{i}({\bf k})-\int\frac{d{\bf k}}{\sqrt{2\omega}}[\tilde J^{i*}({\bf k},t)a^{i}({\bf k})+\tilde J^{i}({\bf k},t)a^{i\dagger}({\bf k})],
\end{eqnarray}
\noindent where
\begin{subequations}
\begin{eqnarray}\label{B2a}
\hat A^{i}({\bf x}) &=&\frac{1}{(2\pi)^{3/2}} \int d{\bf k}\frac{1}{\sqrt{2\omega}}\Bigg[a^{i}({\bf k})e^{i{\bf k}\cdot{\bf x}}+a^{i\dagger}({\bf k})e^{-i{\bf k}\cdot{\bf x}}\Bigg],\\   \label{B2b}                 
 \tilde J^{i}({\bf k,t})&\equiv&\frac{1}{(2\pi)^{3/2}}\int d{\bf x}J^{i}({\bf x},t)e^{-i{\bf k}\cdot{\bf x}}.
\end{eqnarray}
\end{subequations}
 \noindent We shall look for a solution of Schr\"odinger's equation in the form of a coherent state:
 \begin{equation}\label{B3}
|\psi_{A},t\rangle =e^{\int d{\bf k}\alpha^{i}({\bf k},t)a^{i\dagger}({\bf k})}|0\rangle e^{c(t)}.
\end{equation}

 Inserting (\ref{B3}) into Schr\"odinger's equation with the Hamiltonian (\ref{B1}), and utilizing $a^{i}({\bf k})|\psi,t\rangle=\alpha^{i}({\bf k},t)|\psi,t\rangle$, we obtain the two equations:
 \begin{subequations}
 \begin{eqnarray}\label{B4a}
i\dot\alpha^{i}({\bf k},t)&=&\omega \alpha^{i}({\bf k},t)-\frac{1}{\sqrt{2\omega}}\tilde J^{i}({\bf k},t),\\\label{B4b}
i\dot c(t)&=&-\int d{\bf k}\frac{1}{\sqrt{2\omega}}\tilde J^{i*}({\bf k},t)\alpha^{i}({\bf k},t),
\end{eqnarray}
\end{subequations}
\noindent with solutions:
 \begin{subequations}
\begin{eqnarray}\label{B5a}
\alpha^{i}({\bf k},t)&=&i \frac{1}{\sqrt{2\omega}}\int_{0}^{t}dt 'e^{-i\omega(t-t')}\tilde J^{i}({\bf k},t')\nonumber\\
                &=&\frac{1}{(2\pi)^{3/2}}\int d{\bf x}e^{-i{\bf k}\cdot{\bf x}}\Bigg[\sqrt{\frac{\omega}{2}} A_{cl}^{i}({\bf x},t)+i\frac{1}{\sqrt{2\omega}} \dot A_{cl}^{i}({\bf x},t)\Bigg]  \\                                             
c(t)&=&-\int \frac{d{\bf k}}{2\omega}\int_{0}^{t}dt'\int_{0}^{t'}dt''e^{-i\omega(t'-t'')}\tilde J^{i*}({\bf k},t')\tilde J^{i}({\bf k},t'').\label{B5b}\nonumber\\
&=&\frac{i}{2}\int_{0}^{t}dt' \int d{\bf x}A^{i}_{cl}({\bf x},t')J^{i}({\bf x},t')-\frac{1}{2}\int d{\bf k}|\alpha^{i}({\bf k},t)|^{2},
\end{eqnarray}
\end{subequations}
\noindent where we have used (\ref{4b}) with $c^{i}({\bf k})=0$, i.e., for simplicity, we have assumed that all classical currents initially vanish, so $|0\rangle$ is the initial state. Thus, the solenoid current must be established first, and enough time let lapse for the constant vector potential in the neighborhood of the electron to be set up, before the electron moves.  

To prove that this solution is identical to the previously obtained wave function for the vector potential (\ref{A13}).  we note that (\ref{B3}) is an over-complete set of vectors if we regard 
$A_{cl}^{i}({\bf x},t)$ and $\dot A_{cl}^{i}({\bf x},t)$ in (\ref{B5a}) abstractly, as arbitrarily choosable functions.  Thus, our proof is complete if we show that the scalar product of any two of these vectors calculated using (\ref{B3}) is the same as the scalar product of those vectors using (\ref{A13}).  

Denoting the two vectors by the subscripts $R$ and $L$. we have:
\begin{eqnarray}\label{B6}
_{L}\langle\psi_{A},t|\psi_{A},t\rangle_{R}&=&e^{\int d{\bf k}\alpha_{L}^{i*}({\bf k},t)\alpha_{R}^{i}({\bf k},t)}e^{c_{R}^{*}(t)+c_{L}(t)}\nonumber\\
&=&e^{\int d{\bf k} \frac{1}{(2\pi)^{3/2}}\int d{\bf x}'e^{i{\bf k}\cdot{\bf x}'}[\sqrt{\frac{\omega}{2}} A_{L}^{i}({\bf x}',t)-i\frac{1}{\sqrt{2\omega}} \dot A_{L}^{i}({\bf x}',t)]         
\frac{1}{(2\pi)^{3/2}}\int d{\bf x}e^{-i{\bf k}\cdot{\bf x}}[\sqrt{\frac{\omega}{2}} A_{R}^{i}({\bf x},t)+i\frac{1}{\sqrt{2\omega}} \dot A_{R}^{i}({\bf x},t)]}\nonumber\\
&\cdot&e^{-\frac{1}{2}\int d{\bf k} \frac{1}{(2\pi)^{3/2}}\int d{\bf x}'e^{i{\bf k}\cdot{\bf x}'}[\sqrt{\frac{\omega}{2}} A_{R}^{i}({\bf x}',t)-i\frac{1}{\sqrt{2\omega}} \dot A_{R}^{i}({\bf x}',t)]         
\frac{1}{(2\pi)^{3/2}}\int d{\bf x}e^{-i{\bf k}\cdot{\bf x}}[\sqrt{\frac{\omega}{2}} A_{R}^{i}({\bf x},t)+i\frac{1}{\sqrt{2\omega}} \dot A_{R}^{i}({\bf x},t)]}\nonumber\\
&\cdot&e^{-\frac{1}{2}\int d{\bf k} \frac{1}{(2\pi)^{3/2}}\int d{\bf x}'e^{i{\bf k}\cdot{\bf x}'}[\sqrt{\frac{\omega}{2}} A_{L}^{i}({\bf x}',t)-i\frac{1}{\sqrt{2\omega}} \dot A_{L}^{i}({\bf x}',t)]         
\frac{1}{(2\pi)^{3/2}}\int d{\bf x}e^{-i{\bf k}\cdot{\bf x}}[\sqrt{\frac{\omega}{2}} A_{L}^{i}({\bf x},t)+i\frac{1}{\sqrt{2\omega}} \dot A_{L}^{i}({\bf x},t)]}\nonumber\\
&\cdot&e^{  \frac{i}{2}\int_{0}^{t}dt' \int d{\bf x}A^{i}_{R}({\bf x},t')J_{R}^{i}({\bf x},t')-\frac{i}{2}\int_{0}^{t}dt' \int d{\bf x}A^{i}_{L}({\bf x},t')J{L}^{i}({\bf x},t')}\nonumber\\
&=&e^{-\frac{1}{4}\int d{\bf x}'\int d{\bf x}[ A_{L}^{i}({\bf x}',t)-A_{R}^{i}({\bf x}',t)][ A_{L}^{i}({\bf x},t)-A_{R}^{i}({\bf x},t)]\frac{1}{(2\pi)^{3}}\int d{\bf k}\omega 
e^{i{\bf k}\cdot[{\bf x}-{\bf x}']}}\nonumber\\
&&\cdot e^{-\frac{1}{4}\int d{\bf x}'\int d{\bf x}[ \dot A_{L}^{i}({\bf x}',t)-\dot A_{R}^{i}({\bf x}',t)][ \dot A_{L}^{i}({\bf x},t)-\dot A_{R}^{i}({\bf x},t)]\frac{1}{(2\pi)^{3}}\int d{\bf k} \omega^{-1}
e^{i{\bf k}\cdot[{\bf x}-{\bf x}']}}\nonumber\\
&&\cdot e^{\frac{i}{2}\int d{\bf x}[ A_{L}^{i}({\bf x},t)\dot A_{R}^{i}({\bf x},t)- A_{R}^{i}({\bf x},t)\dot A_{L}^{i}({\bf x},t) ]  }
e^{  \frac{i}{2}\int_{0}^{t}dt' \int d{\bf x}[A^{i}_{R}({\bf x},t')J_{R}^{i}({\bf x},t')-A^{i}_{L}({\bf x},t')J_{L}^{i}({\bf x},t')]}.
\end{eqnarray}

On the other hand, from (\ref{A15}) we have:

\begin{eqnarray}\label{B7}
_{L}\langle\psi_{A},t|\psi_{A},t\rangle_{R}&=&\int DA_{L}\langle\psi_{A},t|\psi_{A},t\rangle_{R}\nonumber\\
&=&
\int DA e^{-\int d{\bf x} d{\bf x}'[A^{i}({\bf x})-A_{R}^{i}({\bf x},t) ]\frac{1}{2(2\pi)^{3}}\int d{\bf k}\omega e^{i{\bf k}\cdot ({\bf x}-{\bf x}')}[A^{i}({\bf x}')-
A_{R,cl}^{i}({\bf x}',t)]}e^{i\int d{\bf x}'\dot A_{R}^{i}({\bf x},t)[A^{i}({\bf x})-A_{R}^{i}({\bf x},t) ]}\nonumber\\
&&\cdot e^{-\int d{\bf x} d{\bf x}'[A^{i}({\bf x})-A_{L}^{i}({\bf x},t) ]\frac{1}{2(2\pi)^{3}}\int d{\bf k}\omega e^{i{\bf k}\cdot ({\bf x}-{\bf x}')}[A^{i}({\bf x}')-
A_{L}^{i}({\bf x}',t)]}e^{-i\int d{\bf x}'\dot A_{L}^{i}({\bf x},t)[A^{i}({\bf x})-A_{L}^{i}({\bf x},t) ]}\nonumber\\
&&\cdot e^{\frac{i}{2}\int d{\bf x}[\dot A_{R}^{i}({\bf x},t)A_{R}^{i}({\bf x},t)-\dot A_{L}^{i}({\bf x},t)A_{L}^{i}({\bf x},t] +
\frac{i}{2}\int_{0}^{t}dt'\int d{\bf x}[A_{R}^{i}({\bf x},t')J_{R}^{i}({\bf x},t')- A_{L}^{i}({\bf x},t')J_{L}^{i}({\bf x},t')               ]}.
\end{eqnarray}

We can apply
\begin{equation}\label{B8}
\sqrt{\frac{2}{\pi}}\int_{-\infty}^{\infty}dx
e^{-(x-a)^{2}}e^{-(x-b)^{2}}e^{ip^{1}(x-a)}e^{-ip^{2}(x-b)}
=e^{-\frac{(a-b)^{2}}{2} }           e^{-\frac{(p^{1}-p^{2})^{2}}{8}}  e^{-i\frac{(p^{1}+p^{2})(a-b)}{2}} 
\end{equation}
\noindent to Eq.(\ref{B7}) by diagonalizing its gaussian exponents.  The result is
\begin{eqnarray}\label{B9}
_{L}\langle\psi_{A},t|\psi_{A},t\rangle_{R}&=&e^{-\frac{1}{2}\int d{\bf x} d{\bf x}'[A_{R}^{i}({\bf x},t)-A_{L}^{i}({\bf x},t) ]B({\bf x}-{\bf x}')[A_{R}^{i}({\bf x}',t)-A_{L}^{i}({\bf x}',t) ]}
 e^{-\frac{1}{8}\int d{\bf x} d{\bf x}'[\dot A_{R}^{i}({\bf x},t)-\dot A_{L}^{i}({\bf x},t) ]B^{-1}({\bf x}-{\bf x}')[\dot A_{R}^{i}({\bf x}',t)-\dot A_{L}^{i}({\bf x}',t) ]}\nonumber\\
&&\qquad\qquad\cdot e^{-\frac{i}{2}\int d{\bf x}[\dot A_{R,cl}^{i}({\bf x},t)+\dot A_{L,cl}^{i}({\bf x},t) ][A_{R,cl}^{i}({\bf x},t)- A_{L,cl}^{i}({\bf x},t) ]}\nonumber\\
&&\cdot e^{\frac{i}{2}\int d{\bf x}[\dot A_{R}^{i}({\bf x},t)A_{R}^{i}({\bf x},t)-\dot A_{L}^{i}({\bf x},t)A_{L}^{i}({\bf x},t] +
\frac{i}{2}\int_{0}^{t}dt'\int d{\bf x}[A_{R}^{i}({\bf x},t')J_{R}^{i}({\bf x},t')- A_{L}^{i}({\bf x},t')J_{L}^{i}({\bf x},t')               ]}.
 \end{eqnarray}
\noindent where $B({\bf x}-{\bf x}')$ is given by  (\ref{A6})  and  $B^{-1}({\bf x}-{\bf x}')\equiv \frac{2}{(2\pi)^{3}}\int d{\bf k}\omega^{-1} e^{i{\bf k}\cdot ({\bf x}-{\bf x}')}$.  

Combining terms on the second and third lines of (\ref{B9}), we see that expressions (\ref{B6}) and (\ref{B9}) are identical.

\section{Overlap Integral}

We discuss here the contribution to the overlap integral $\langle \psi_{A,L},t|\psi_{A,R},t\rangle$ of the first line in Eq.(\ref{6}). The amplitude and phase are given by the first two lines of Eq. (\ref{B9}).
\subsection{The Phase}
The second line of  (\ref{B9}) contains the phase

\begin{equation}\label{C1}
-\frac{1}{2}\int d{\bf x}[\dot A_{R,el}^{i}({\bf x},t)+\dot A_{L,el}^{i}({\bf x},t) ][A_{R,el}^{i}({\bf x},t)- A_{L,el}^{i}({\bf x},t) ].
\end{equation}
\noindent Because the left and right traverses have the same constant solenoid vector potential, $A_{R,sol}^{i}({\bf x})= A_{Lsol}^{i}({\bf x},t)$, this phase  has been expressed in terms of the electron's classical vector potential  alone. Under rotation of the physical situation about the $y$-axis by 180$^{\circ}$, $A_{L,el}^{i}({\bf x},t)\leftrightarrow A_{R,el}^{i}({\bf x},t) $, and so the integrand of (\ref{C1}) changes sign.  Since this is achieved by a  coordinate transformation, Eq.(\ref{C1}) is equal to its negative, and therefore vanishes. Therefore, there is no phase shift contribution by the first term in  (\ref{6}).

\subsection{The Amplitude: Physical Nature}
Now, consider the amplitude of the overlap integral $\equiv e^{-a(t)}$, given by the first line in (\ref{B9}):
\begin{eqnarray}\label{C2}
a(t)&=&\frac{1}{4}\int d{\bf x} d{\bf x}'A_{RL}^{i}({\bf x},t)\frac{1}{(2\pi)^{3}}\int d{\bf k}\omega e^{i{\bf k}\cdot ({\bf x}-{\bf x}')}A_{RL}^{i}({\bf x}',t)
\nonumber\\
&&\qquad\qquad+\frac{1}{4}\int d{\bf x} d{\bf x}'\dot A_{RL}^{i}({\bf x},t)\frac{1}{(2\pi)^{3}}\int d{\bf k}\omega^{-1} e^{i{\bf k}\cdot ({\bf x}-{\bf x}')}\dot A_{RL}^{i}({\bf x}',t) 
\end{eqnarray}
\noindent where $A_{RL}^{i}({\bf x},t)\equiv A_{R}^{i}({\bf x},t)-A_{L}^{i}({\bf x},t).$  Suppose that $A_{RL}^{i}({\bf x},t)$ is generated by a current for $0\leq t\leq T$, but the current vanishes for $t>T$, so the field generated during $0\leq t\leq T$ propagates freely for $t>T$.  Then, we can show that $a(t)$ is a constant of the motion for $t>T$, as follows:
\begin{eqnarray}\label{C3}
\frac{d}{dt}a(t)&=&\frac{1}{2}\int d{\bf x} d{\bf x}'\dot A_{RL}^{i}({\bf x},t)\frac{1}{(2\pi)^{3}}\int d{\bf k}\omega e^{i{\bf k}\cdot ({\bf x}-{\bf x}')}A_{RL}^{i}({\bf x}',t)
\nonumber\\
&&\qquad\qquad+\frac{1}{2}\int d{\bf x} d{\bf x}'\ddot A_{RL}^{i}({\bf x},t)\frac{1}{(2\pi)^{3}}\int d{\bf k}\omega^{-1} e^{i{\bf k}\cdot ({\bf x}-{\bf x}')}\dot A_{RL}^{i}({\bf x}',t) \nonumber\\
&=&\frac{1}{2}\int d{\bf x} d{\bf x}'\dot A_{RL}^{i}({\bf x},t)\frac{1}{(2\pi)^{3}}\int d{\bf k}\omega e^{i{\bf k}\cdot ({\bf x}-{\bf x}')}A_{RL}^{i}({\bf x}',t)
\nonumber\\
&&\qquad\qquad+\frac{1}{2}\int d{\bf x} d{\bf x}'\nabla^{2} A_{RL}^{i}({\bf x},t)\frac{1}{(2\pi)^{3}}\int d{\bf k}\omega^{-1} e^{i{\bf k}\cdot ({\bf x}-{\bf x}')}\dot A_{RL}^{i}({\bf x}',t) \nonumber\\
&=&\frac{1}{2}\int d{\bf x} d{\bf x}'\dot A_{RL}^{i}({\bf x},t)\frac{1}{(2\pi)^{3}}\int d{\bf k}\omega e^{i{\bf k}\cdot ({\bf x}-{\bf x}')}A_{RL}^{i}({\bf x}',t)
\nonumber\\
&&\qquad\qquad-\frac{1}{2}\int d{\bf x} d{\bf x}' A_{RL}^{i}({\bf x},t)\frac{1}{(2\pi)^{3}}\int d{\bf k}\omega e^{i{\bf k}\cdot ({\bf x}-{\bf x}')}\dot A_{RL}^{i}({\bf x}',t)=0
\end{eqnarray}
\noindent where, in going from the second line to the fourth line, the free propagation of $A_{RL}^{i}({\bf x},t)$ has been utilized. 

What is this constant of the motion?  Writing the free field in terms of its Fourier components as
\[ 
A_{RL}^{i}({\bf x},t)=\frac{1}{(2\pi)^{3/2}}\int d{\bf k}\frac{1}{\sqrt{2\omega}}\Big[a^{i}(k)e^{i{\bf k}\cdot{\bf x}-i\omega t}+a^{i*}(k)e^{-i{\bf k}\cdot{\bf x}+i\omega t}\Big]
\]
\noindent we find
\[
a(t)=\int d{\bf k}a^{i*}({\bf k})a^{i}({\bf k}).
\]
\noindent Were $a^{i}({\bf k}),a^{i*}({\bf k})$ annihilation and  creation operators instead of c-number amplitudes, $a(t)$ would be the photon number operator. So, we may think of $a(t)$ as   a classical analog of the difference of the number of photons for the left and right traverses.  

\subsection{The Amplitude: Divergence}

Expressing the vector potential in terms of the current using the second line of (\ref{4b}), performing the integrals over ${\bf x},{\bf x}'$ and then over the delta functions, we obtain
\begin{eqnarray}\label{C4}
a(t)&=&\frac{1}{4}\frac{1}{(2\pi)^{3}}\int_{0}^{t}dt_{1}\int_{0}^{t}dt_{2}\int d{\bf x}_{1}d {\bf x}_{2}\int d{\bf k}e^{i{\bf k}\cdot({\bf x}_{1}- {\bf x}_{2})}\frac{1}{k}\cos k(t_{1}-t_{2})\nonumber\\
&&\qquad\qquad\qquad[{\bf J}_{L}({\bf x}_{1},t_{1})-{\bf J}_{R}({\bf x}_{1},t_{1})]\cdot[{\bf J}_{L}({\bf x}_{2},t_{2})-{\bf J}_{R}({\bf x}_{2},t_{2})]
\end{eqnarray}
\noindent Then, writing ${\bf k}\cdot({\bf x}_{1}- {\bf x}_{2})= k|{\bf x}_{1}- {\bf x}_{2}|\cos\theta$, integrating over $\theta$, writing $\sin k|{\bf x}_{1}- {\bf x}_{2}|\cos k(t_{1}-t_{2})$ as 
the sum of $\sin$'s, and using $\int_{0}^{\infty} dk\sin kz={\cal P}\frac{1}{z}$, where ${\cal P}$ is the principal part, we obtain
\begin{eqnarray}\label{C5}
a(t)&=&\frac{1}{4}\frac{1}{(2\pi)^{2}}\int_{0}^{t}dt_{1}\int_{0}^{t}dt_{2}\int d{\bf x}_{1}d {\bf x}_{2}\frac{1}{|{\bf x}_{1}- {\bf x}_{2}|}{\cal P}\Big[\frac{1}{|{\bf x}_{1}- {\bf x}_{2}|+(t_{1}-t_{2})}+
\frac{1}{|{\bf x}_{1}- {\bf x}_{2}|-(t_{1}-t_{2})}\Big]\nonumber\\
&&\qquad\qquad\qquad[{\bf J}_{L}({\bf x}_{1},t_{1})-{\bf J}_{R}({\bf x}_{1},t_{1})]\cdot[{\bf J}_{L}({\bf x}_{2},t_{2})-{\bf J}_{R}({\bf x}_{2},t_{2})].
\end{eqnarray}

Now,  a classical electron orbiting counterclockwise in a half-circle of radius R with speed $u$ in the $z=0$ plane, starting at $\phi=-\pi/2$ at time 0, and ending at 
$\phi=\pi/2$ at time $T=\pi R/u$ provides the current ${\bf J}_{R}({\bf x},t)=e{\bf u}_{R}(t)\delta({\bf x}-{\bf R}_{R}(t))$, where ${\bf u}_{R}(t)=u[-{\bf i}\sin\phi_{R}(t)+{\bf j}\cos\phi_{R}(t)],$ ${\bf R}_{R}(t)=R[{\bf i}\cos\phi_{R}(t)+{\bf j}\sin\phi_{R}(t)]$, $\phi_{R}(t)=-\frac{\pi}{2}+\frac{ut}{R}$.

Similarly, the electron orbiting clockwise starting at  $\phi=3\pi/2$ (this is identified with $\phi=\pi/2$, so that the discontinuity in angle is at $x=0, y=-R$) at time 0 and ending also at 
$\phi=\pi/2$ at time $T=\pi R/u$ provides the current ${\bf J}_{L}({\bf x},t)=e{\bf u}_{L}(t)\delta({\bf x}-{\bf R}_{L}(t))$, where 
 ${\bf u}_{L}(t)=u[-{\bf i}\sin\phi_{L}(t)+{\bf j}\cos\phi_{L}(t)]$, ${\bf R}_{L}(t)=R[{\bf i}\cos\phi_{L}(t)+{\bf j}\sin\phi_{L}(t)]$, $\phi_{L}(t)=\frac{3\pi}{2}-\frac{ut}{R}$.
  Putting these currents into (\ref{C5}) results in:
 \begin{eqnarray}\label{C6}
a(t)&=&e^{2}u^{2}\frac{1}{(2\pi)^{2}}\int_{0}^{t}dt_{1}\int_{0}^{t}dt_{2}{\cal P}
\Big[\frac{\cos\frac{u}{R}(t_{1}-t_{2})}{4R^{2}\sin^{2}\frac{u}{2R}(t_{1}-t_{2}) - (t_{1}-t_{2})^{2} } -  \frac{\cos\frac{u}{R}(t_{1}+t_{2})}{4R^{2}\sin^{2}\frac{u}{2R}(t_{1}+t_{2}) - (t_{1}-t_{2})^{2} } \Big]
\end{eqnarray}

The integral of the first term in the bracket over $(t_{1}-t_{2})$ is divergent. 

\subsection{Removing the Divergence: Smearing the Charge}

The divergence is due to the ``self-interacting" terms  in (\ref{C4}) or (\ref{C5}), ${\bf J}_{L}({\bf x}_{1},t_{1}){\bf J}_{L}({\bf x}_{2},t_{2})$ and  ${\bf J}_{R}({\bf x}_{1},t_{1}){\bf J}_{R}({\bf x}_{2},t_{2})$.  It occurs when $t_{1}=t_{2}$, when the point electron is superimposed upon itself.  This suggests that the divergence might be ameliorated by smearing out the charge. 

 For example, if the current was due to a  uniform ``ball" of charge, we would set ${\bf J}({\bf x},t)=e{\bf u}_{R}(t)\int d{\bf x}' \rho({\bf x}')(\delta({\bf x}-{\bf R}_{R}(t)-{\bf x}')$ with $\rho({\bf x}')=\frac{\Theta(\sigma-|{\bf x}'|)}{4\pi\sigma^{3}/3}$.  
 
 We shall adopt the simplest ``smearing," extending the point charge into a line charge in the $z$-direction of length $\sigma,$ setting 
 $\rho({\bf x}')=\delta(x')\delta(y')\Theta(\sigma-z')\Theta(z')/\sigma$.  Then, 
 Eq.(\ref{C6}) becomes, after the electron packets complete their traverse at time $T$,  utilizing $uT=\pi R$ and putting in $c$ and $\hbar$:
\begin{eqnarray}\label{C7}
a(T)&=&\frac{e^{2}}{\hbar c}\frac{u^{2}}{4\sigma^{2}}\frac{1}{(2\pi)^{2}}\int_{0}^{\sigma} dz'_{1}\int_{0}^{\sigma}dz'_{2}\int_{0}^{T}dt_{1}\int_{0}^{T}dt_{2}\nonumber\\
&&{\cal P}
\Bigg[\frac{\cos\frac{\pi}{T}(t_{1}-t_{2})}{4(uT/\pi)^{2}\sin^{2}\frac{\pi}{2T}(t_{1}-t_{2})+(z'_{1}-z'_{2})^{2}- c^{2}(t_{1}-t_{2})^{2} } -  
\frac{\cos\frac{\pi}{T}(t_{1}+t_{2})}{4(uT/\pi)^{2}\sin^{2}\frac{\pi}{2T}(t_{1}+t_{2})+(z'_{1}-z'_{2})^{2} - c^{2}(t_{1}-t_{2})^{2} } \Bigg]\nonumber\\
&=&\frac{e^{2}}{\hbar c}\frac{\beta^{2}}{4}\frac{1}{(2\pi)^{2}}\int_{0}^{1} dz_{1}\int_{0}^{1}dz_{2}\int_{0}^{1}d\tau_{1}\int_{0}^{1}d\tau_{2}\nonumber\\
&&{\cal P}
\Bigg[\frac{\cos\pi(\tau_{1}-\tau_{2})}{(\beta/\pi)^{2}[4\sin^{2}\frac{\pi}{2}(\tau_{1}-\tau_{2})+\lambda^{2}(z_{1}-z_{2})^{2}]-(\tau_{1}-\tau_{2})^{2} } -  
\frac{\cos\pi(\tau_{1}+\tau_{2})}{(\beta/\pi)^{2}[4\sin^{2}\frac{\pi}{2}(\tau_{1}+\tau_{2})+\lambda^{2}(z_{1}-z_{2})^{2}]-(\tau_{1}-\tau_{2})^{2} }\Bigg] \nonumber\\
\end{eqnarray}
\noindent where in the second expression we have changed to dimensionless variables $\tau_{i}\equiv t_{i}/T$, $z_{i}\equiv z'_{i}/\sigma$, $\beta\equiv u/c$ and
 $\lambda\equiv(\sigma/R)$.  
 
 Now, we focus on the divergent first term in the bracket of (\ref{C7}) which we shall call $a_{1}(T)$.  Change variables to $\tau_{\pm}\equiv \tau_{1}\pm\tau_{2}$, $z_{\pm}\equiv z_{1}\pm z_{2}$, and use 
 \[
\int_{0}^{1}d \tau_{1}\int_{0}^{1}d \tau_{2}= \frac{1}{2}\Big[\int_{-1}^{0}d \tau_{-} \int_{-\tau_{-}}^{2+\tau_{-}}d \tau_{+} +\int_{0}^{1}d \tau_{-} \int_{\tau_{-}}^{2-\tau_{-}}d \tau_{+} \Big] 
=2\int_{0}^{1}d \tau_{-}[1- \tau_{-}],       
\]
and similarly for $z_{\pm}$, since the integral does not depend upon $\tau_{+}, z_{+}$, and depends upon the square of each of $\tau_{-}, z_{-}$.  $a_{1}(T)$ is then
\begin{eqnarray}\label{C8}
a_{1}(T)&=&
\frac{e^{2}}{\hbar c}\frac{\beta^{2}}{4}\frac{1}{(2\pi)^{2}}\int_{0}^{1}d\tau_{-}[1-\tau_{-}] \cos\pi\tau_{-}\int_{0}^{1}dz_{-}[1- z_{-}]  
{\cal P}
\frac{1}{                      (\beta\lambda/\pi)^{2}z_{-}^{2} - [\tau_{-}^{2} -   (2\beta/\pi)^{2}\sin^{2}\frac{\pi}{2}\tau_{-}] }.
\end{eqnarray}
\noindent We may neglect $(2\beta/\pi)^{2}\sin^{2}\frac{\pi}{2}\tau_{-}$ compared to $\tau_{-}^{2}$, since it is a factor  $\beta^{2}$ smaller. 

 For $|\alpha|<1$, we have
 \begin{subequations}
\begin{eqnarray}\label{C9}
{\cal P} \int_{0}^{1}dz\frac{1}{z^{2}-\alpha^{2}}&=&\lim_{\epsilon\rightarrow 0}\Bigg[-\int_{0}^{|\alpha|-\epsilon}dz\frac{1}{\alpha^{2}-z^{2}}+\int_{|\alpha|+\epsilon}^{1}dz\frac{1}{z^{2}-\alpha^{2}}\Bigg]=\frac{1}{2|\alpha|}\ln\frac{1-|\alpha|}{1+|\alpha|},\\\label{C9a}
{\cal P}\int_{0}^{1}dz\frac{z}{z^{2}-\alpha^{2}}&=&\int_{0}^{1}dz\frac{1}{z+|\alpha|}+|\alpha| {\cal P}\int_{0}^{1}dz\frac{1}{z^{2}-\alpha^{2}} =
\ln [1+|\alpha|]-\ln |\alpha|+\frac{1}{2}\ln\frac{1-|\alpha|}{1+|\alpha|}.
 \label{C9b}  
\end{eqnarray}
 \end{subequations}

Since $|\alpha|=\frac{\pi}{\beta\lambda}|\tau_{-}|,$ it is clear that the subsequent integral $\int_{0}^{1}d\tau_{-}=\frac{\beta\lambda}{\pi}\int_{0}^{\frac{\pi}{\beta\lambda}}d|\alpha|$ no longer diverges.  

The  integral in (\ref{C8})  is $\sim\frac{1}{\beta\lambda}$,  so $ a_{1}(T)\sim \frac{e^{2}}{\hbar c}\frac{u}{c}\frac{R}{\sigma}$. This result is used in Section III.  

The second term in the bracket of (\ref{C7}), which we shall call $a_{2}(T)$, is not divergent and does not need smearing, so we may set $\lambda=0$.  
$a_{2}(T)$ is a factor $\beta^{2}$ smaller than $ a_{1}(T)$.

\acknowledgements{ We would like to thank Lev Vaidman for very helpful conversations. }


\begin{thebibliography}{99}

 \bibitem{Ton} Akira Tonomura, Nobuyuki Osakabe, Tsuyoshi Matsuda, Takeshi Kawasaki, and Junji Endo, Phys. Rev. Lett. vol. 56, pp. 792Ð795 (1986); Osakabe, N; et al.  Physical Review A 34  815Ð822 (1986). 

\bibitem{AB} Y. Aharonov and D. Bohm, Phys. Rev. {\bf 115}, 485-491 (1959),  Phys. Rev. 123, 1511 (1961) .

\bibitem{FR} W. H. Furry, N. F. Ramsey, Rev. {\bf 118}, 623 (1960).

\bibitem{AR} For a nice presentation of the A-B electric and magnetic effects, see Y. Aharonov and D. Rohrlich, \textit{Quantum Paradoxes} (Wiley, Weinheim 2005), Chapter 4. 

\bibitem{Vaidman} L. Vaidman,  Phys. Rev.  A{\bf 86}, 040101 (2012). 

\bibitem{PR} P. Pearle and A. Rizzi,``Quantum Mechanical Inclusion of the Source in the Aharonov-Bohm Effects."





\end{thebibliography}
 \end{document}